\documentclass[aps, prb, twocolumn, superscriptaddress, preprintnumbers, amsmath, epsFig, floatfix,normalem]{revtex4}
\usepackage{graphicx}
\usepackage{dcolumn}
\usepackage{bm}
\usepackage{ulem}
\usepackage{amsmath}
\usepackage{textcomp}
\usepackage{amssymb}
\usepackage[utf8]{inputenc}
\usepackage[english]{babel}
\usepackage{keyval}
\graphicspath{{Figs/}{Figs:}{\Figs}}
\setkeys{Gin}{width=0.9\columnwidth}

\usepackage[usenames,dvipsnames]{color}
\usepackage[]{ulem}

\newcommand{\comment}[1]{\textcolor{red}{#1}}
\renewcommand{\comment}[1]{\relax}

\newcommand{\todelete}[1]{\textcolor{green}{\sout{#1}}}
\renewcommand{\todelete}[1]{\relax}

\begin{document}

\title{Observation of Shubnikov-de Haas Oscillations, Non-trivial Berry Phase, Planar Hall and Anisotropic Magnetoresistance at the conducting interface of EuO-KTaO$_3$}
\date{\today}

\author{Nand Kumar}
\affiliation{Nanoscale Physics and Device Laboratory, Institute of Nano Science and Technology, Phase- 10, Sector- 64 Mohali, Punjab - 160062, India.}
\affiliation{Applied Science Department,Punjab Engineering College (Deemed to be University), Sector-12, Chandigarh - 160012, India.}

\author{Neha Wadehra}
\affiliation{Nanoscale Physics and Device Laboratory, Institute of Nano Science and Technology, Phase- 10, Sector- 64 Mohali, Punjab - 160062, India.}

\author{Ruchi Tomar}
\affiliation{Nanoscale Physics and Device Laboratory, Institute of Nano Science and Technology, Phase- 10, Sector- 64 Mohali, Punjab - 160062, India.}

\author{Sushanta Dattagupta}
\affiliation{Bose Institute, P-1/12, CIT Rd, Scheme VIIM, Kankuragachi, Kolkata, West Bengal-700054, India.} 

\author{Sanjeev Kumar}
\affiliation{Applied Science Department,Punjab Engineering College (Deemed to be University), Sector-12, Chandigarh - 160012, India.}

\author{S. Chakraverty}
\email{suvankar.chakraverty@inst.ac.in}
\affiliation{Nanoscale Physics and Device Laboratory, Institute of Nano Science and Technology, Phase- 10, Sector- 64 Mohali, Punjab - 160062, India.}

\begin{abstract}
\noindent
The momentum dependent splitting of spin-bands in an electronic system is known as the "Rashba effect". Systems with the "Rashba effect" possess a Dirac point in momentum space. An electron in a cyclotron orbit enclosing that Dirac point in the reciprocal space gains a "Berry phase". We report here  the Shubnikov-de-Haas oscillations (SdH) at the conducting interface of EuO-KTaO$_3$ (KTO). Observed SdH oscillations suggest the presence of two Fermi surfaces. For both the Fermi surfaces, we have seen the presence of a non-trivial "Berry phase" suggesting that the surfaces enclose the "Dirac point". Thus the Berry phase originates from the inner and outer Fermi surfaces of the Rashba spin-split bands. As in topological insulators, two fold planar Hall and anisotropic magnetoresistance have also been observed in EuO-KTO. Analyzing the SdH, Hall and magnetoresistance data, we have drawn a possible band diagram near the Fermi surface. 
\end{abstract}

\maketitle
In recent times, manipulation of spin of an electron in addition to its charge has attracted tremendous attention and given rise to a new field called  "spintronics".\cite{Datta1990,Meier2002} Segregation of spin-bands of the charge carriers is one of the basic principles of spintronic devices. One of the ways to achieve spin-splitting is the "Rashba effect". \cite{Bychkov1984} It exploits the pseudo magnetic field originating from the broken spatial symmetry of a system at the frame of reference of relativistic electrons, yielding a momentum-dependent spin-splitting.\cite{Ishizaka2011} An accompanying feature of the Rashba Effect is the presence of a strong spin-orit coupling (SOC). Rashba type spin-splitting is interesting not only because of its technological implications but also, such systems can exhibit rich physics arising from the topological aspects of the system.\cite{Inoue2009,Murakawa2013}

There has been a continuous and extensive effort to realize materials exhibiting Rashba effect. Rashba effect has already been observed in semiconductors and topological insulators.\cite{Eisentein1984,Nitta1997,Zhang2010,Bahramy2012} Another important class of materials which attracted recent interest in this regard is "Oxides".\cite{Bibes2011,Shanavas2014} This is manily because "oxides" specially "perovskite oxides" possess a variety of physical properties, which if integrated with Rashba system may exhibit emergent phenomena.\cite{Bibes2011,Tokura1994,Schooley1964,Tikhomirov2002,Lee2013,Wadehra2019} 

It has been reported that conducting interface can be created by juxtaposing a non-polar perovskite oxide with another polar perovskite oxide.\cite{Ohtomo2004,Hotta2007} In such conducting interfaces, an electric field is developed due to the asymmetric potential of the two sides of the interface. Such conducting interface if combined with strong SOC and relativistic conduction electrons, can be a prospective candidate for realizing Rashba effect.\cite{Tomar2018} Most of the oxide based conducting interfaces are realized in SrTiO$_3$ (STO) based systems (such as LaAl$O_3$-STO, LaV$O_3$-STO, LaTi$O_3$-STO etc.).\cite{Mannhart2008} Until recently, conducting interfaces of LaTi$O_3$-KTO and EuO-KTO and were reported.\cite{Zou2015,Zhang2018} KTO based interfaces are even more interesting because all the ingredients of the Rashba systems : SOC, relativistic conduction electrons and polar crystal structure are present.\cite{Wadehra2017,Nakamura2009} Consequences of Rashba effect and related non-trivial Berry phase should be reflected in SdH oscillation.\cite{Murakawa2013} But so far, no SdH oscillations have been reported  for any KTO based conducting interfaces\cite{Zou2015,Zhang2018} to the best of our knowledge.

In this work, we report detection of a non-trivial $\pi$-Berry phase in the conducting heterostructure of EuO-KTO via analysis of SdH oscillations observed in the system.  Our analysis of the SdH oscillation pattern suggests the presence of two Fermi surfaces contributing towards oscillations. The observation of weak-antilocalization (WAL) in our out of plane magnetoresistance (MR) data confirms the presence of high SOC in this conducting heterointerface.\cite{Nakamura2009} The in-plane magnetotransport measurements result in oscillating longitudinal and transverse magnetoresistance, similiar to that observed in topological insulators.\cite{Taskin2017,Rakhmilevich2018} Presence of the WAL in magnetoresistance, observation of two Fermi surfaces and existence of a non-trivial Berry phase for both the Fermi surfaces suggest the Rashba-type spin splitting in EuO-KTO.\cite{Veit2018} Based on our experimental observations, we have drawn a band structure near the Fermi level.

\begin{figure}[t!] \scalebox{1}{\includegraphics{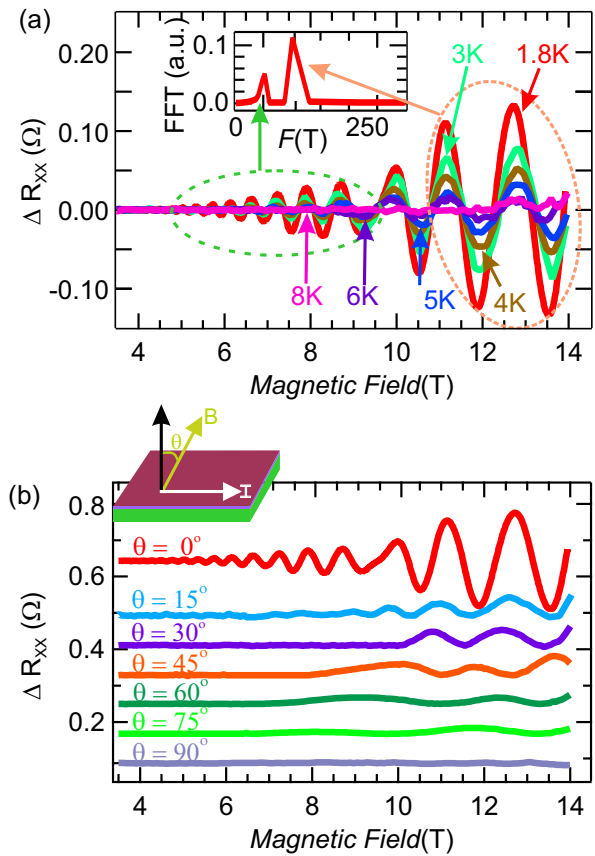}}
\caption{(Color online) (a) $\Delta{R_{xx}}$ as a function of applied magnetic field at different temperatures. Orange and green circles represent high and low magnetic field oscillations respectively. The inset shows the FFT spectrum for oscillations at 1.8 K. (b) Angle dependent $\Delta{R_{xx}}$ as a function of applied magnetic field. The schematic shows the measurement geometry.}
\end{figure}

Conducting interface of EuO-KTO had recently been reported.\cite{Zhang2018} This interface hosts spin polarized conduction electrons. A strong correlation between magnetism and conduction electrons had also been demonstrated through the presence of well defined anomalous Hall effect, that vanishes at the magnetic Curie temperature of EuO. In the current work, we employed pulsed laser deposition technique to grow thin films of ferromagnetic insulator EuO of thickness $\sim$10 nm on insulating (001) oriented single crystals of KTO.\cite{Zhang2018} (The detailed description of growth mechanism and parameters are discussed in supplementary information.) The heterointerface was found to be conducting down to low temperature as shown by the temperature dependent sheet resistivity measurements (supplemetary Fig. S1(b)). Temperature dependent resistivity shows very similar behaviour as reported previously for EuO-KTO interface.\cite{Zhang2018} For the first set of experiments, magnetotransport measurements were performed by applying a magnetic field normal to the sample surface as shown in schematic of Fig. 1(b)($\theta$=0$^o$). Both longitudinal (MR) and transverse (Hall) resistance were measured as a function of the magnetic field. The Hall effect data reveal the presence of an anomalous Hall effect (Fig. S1(c) lower panel), similar to that observed by Zhang et. al.\cite{Zhang2018} The charge carrier density and mobility were calculated from the Hall measurements and were found to be 5.0x10$^{13}$ cm$^{-2}$ and 1380 cm$^2$V$^{-1}$s$^{-1}$ respectively. It is worth noting that the mobility achieved in the present sample is more than two times of that reported by Zhang et. al. Other physical properties, such as carrier density, magnetic Curie temperature, anomalous Hall effect etc. are very similiar to that reported by Zhang et. al.

In the longitudinal geometry, we observed quantum SdH oscillations in the magnetoresistance (R$_{xx}$). The occurrence of the oscillations was further ascertained by subtracting a fifth order polynominal fitting to R$_{xx}$ over a range of the magnetic field from 0 to 14 T. Figure 1(a) shows the $\Delta{R_{xx}}$ as a function of applied magnetic field (B) at different temperatures suggesting that the oscillations disappear above 8 K. The Fast Fourier Transform (FFT) of $\Delta{R_{xx}}$ versus 1/B indicates two frequencies in these oscillations ($\alpha_1$) 90 and ($\alpha_2$) 50 T.\cite{Cancellieri2013} These two oscillations are identified as low (B $\textless$ 10 T) and high field (B $\textgreater$ 10 T) regions and are marked in the Fig. 1(a).\cite{Phillips2014} These two oscillations can simply originate from Rashba spin-split bands. \cite{Murakawa2013,Veit2018,Xiang2015,Ye2015} We then performed angle dependent magnetoresistance measurements by varying the angle between B and the sample surface as shown in schematic of Fig. 1(b). The $\Delta{R_{xx}}$-B curve at different angles ($\theta$) is shown in Fig. 1(b). Figure 1(b) suggests that the low field oscillations (related to $\alpha_2$ =50 T) die out on varying $\theta$ indicating 2D nature of the Fermi surface corresponding to these oscillations.\cite{Xiang2015,Harashima2013,Veit2019,Analytis2010,Ning2014,Wang2011,Kumar2016}  Whereas, the high field oscillations persist even upto $\theta$ = 90$^o$ suggesting 3D nature of the Fermi surface.

\begin{figure*}[t!] \scalebox{2.2}{\includegraphics{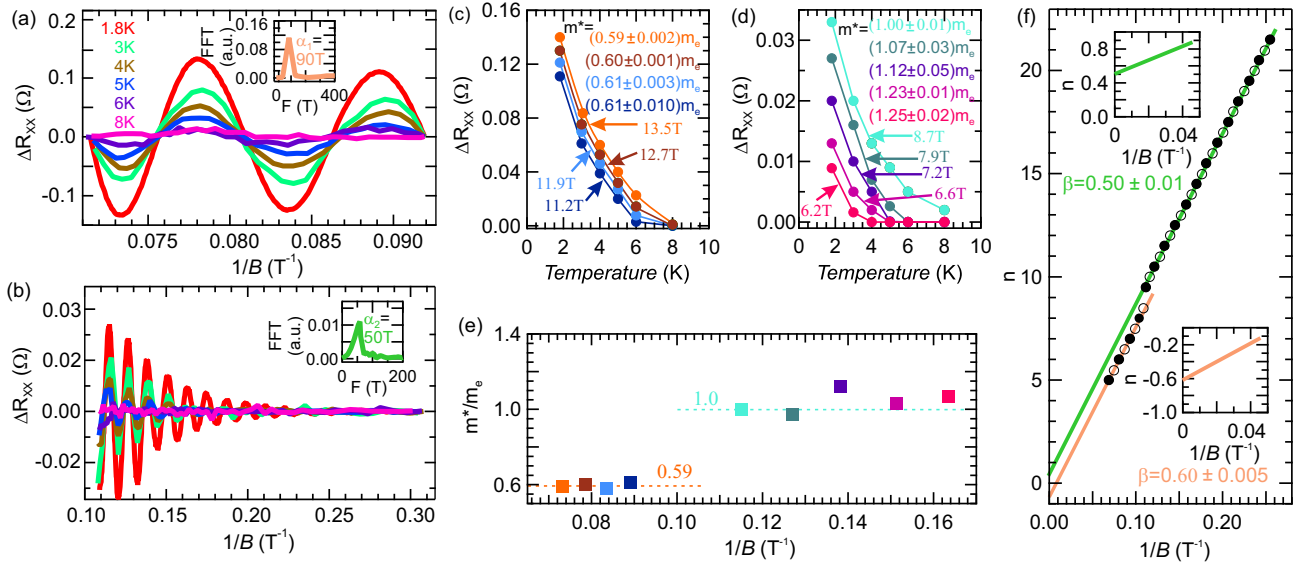}}
\caption{(Color online) (a) and (b) $\Delta{R_{xx}}$ as a function of 1/B at different temperatures for high and low frequency respectively. (c) and (d) Temperature dependence of oscillation amplitudes at 90 and 50 T respectively. The calculated effective masses for different field values at different temperatures are shown. (e) The m*/m$_e$ values as a function of 1/B for low and high fields. (f) Landau index plot for two frequecies. Close and open circles denote the integer ($\Delta{R_{xx}}$, valleys) and half integer ($\Delta{R_{xx}}$, peaks) index respectively. The insets are the zoomed regions around zero.}
\end{figure*}

In order to obtain detailed information about the charge carriers from the SdH oscillations, we isolated and analyzed the two oscillating components of $\Delta{R_{xx}}$ separately. Figures 2(a) and (b) show the temperature dependent quantum oscillations corresponding to frequencies 90 and 50 T. The effective mass (m*) \cite{Shoenberg1984} for both high and low field oscillations was calculated from the temperature dependent oscillation amplitude at different fields (indicated in Fig. 2(c) and (d)) after fitting with the formula derived from Lifshitz-Kosevich theory:\cite{Akiyama2018}

\begin{equation}
\frac{\Delta R_{xx}(T)}{\Delta R_{xx}(0)} = \frac{\lambda(T)}{sinh(\lambda(T))}
\end{equation}

Here, $\lambda(T)$ = 2$\pi^2$m$^*$k$_B$T/$\hbar$eB, m$^*$ is the effective mass, k$_B$ is Boltzman's constant, T is the temperature, e is electric charge and $\hbar$ (1.05x10$^{-34} Js$) is Plank's constant.
Figure 2(e) shows the m*/m$_e$ values as a function of 1/B. This figure suggests that for electrons exhibiting low field oscillations, the effective mass is $\sim$1.0 m$_e$ and that for high field is  $\sim$0.59 m$_e$, where m$_e$ is the mass of free electron.

To obatin the Berry phase, we plotted Landau level fan diagram, a plot of the Landau index (n) as a function 1/B. The oscillations were indexed with valleys as an interger (n) and peaks as half integer (n+1/2).\cite{He2014} Figure 2(f) represents the Landau fan diagram for both types of oscillations. We have defined the close and open circles as the integers ($\Delta{R_{xx}}$, valleys) and half integers ($\Delta{R_{xx}}$, peaks) respectively. This plot was fitted with Lifshitz-Onsagar equation mentioned as:\cite{Murakawa2013}

\begin{equation}
A(\frac{\hbar}{eB})= 2\pi(n+\frac{1}{2}-\frac{\phi_B}{2\pi}+\delta)
\end{equation}

where A is the cross-sectional area of Fermi surface which is related to the Landau level (n), ($\phi_B$/2$\pi$-$\delta$) is the intercept ($\beta$), $\phi_B$ is the Berry phase and $\delta$=0 is for 2D or $\delta$=\textpm 1/8 is for 3D type of the Fermi surface.\cite{Murakawa2013,Zhang2005} The n versus 1/B plot was extrapolated to determine the intercept which gives the Berry phase according to Lifshits-Onsagar quantization rule.\cite{Veit2018,Amit2018} Insets show the magnified view of the graph close to zero. The extrapolation of n versus 1/B clearly represents the intercept ($\phi_B$/2$\pi$ \textpm $\delta$) values to be 0.5 and -0.6 on y-axis for low and high frequencies respectively.

We had already identified that the lower field oscillations have 2D character and hence $\delta$ should be zero. This suggests the electron exhibiting low field SdH oscillations possess a ($\phi_B$/2$\pi$=0.5; $\phi_B$=$\pi$) $\pi$ non-trivial Berry phase.\cite{Amit2018,Xiao2010} Similiarly, electrons exhibiting high field oscillations have 3D character and $\delta$ is \textpm 1/8 for them. These electrons also possess ($\phi_B$/2$\pi$ + 1/8 = 0.6; $\phi_B$=$\pi$) $\pi$ non-trivial Berry phase. These observations suggest that both types of electrons observed in SdH oscillations are orbiting in cyclotron motion enclosing a Dirac point in reciprocal space. Hence, these two conducting channels could be regarded as outer Fermi surface (OFS) and inner Fermi surface (IFS)  of Rashba spin-split band. The area of two Fermi surfaces related to high and low magnetic field are calculated to be 8.3x10$^{-3}$ $\AA^{-2}$ and 4.6x10$^{-3}$ $\AA^{-2}$ respectively.

In addition to this, our low temperature magnetotransport measurements for magnetic field applied perpendicular to the sample surface showed a sharp positive MR, signature of WAL due to strong SOC in the system.\cite{Nakamura2009,Hikami1980,Iordanskii1994} For Rashba-type systems, theory has been developed by Iordanskii, Lyanda-Geller, and Pikus (ILP theory) to describe WAL in magnetoconductance. The expression of the magnetoconductance developed by ILP theory is given as:\cite{Iordanskii1994,Lee2012}

\begin{equation}
\Delta\sigma= \frac{e^2}{2\pi^2\hbar}[ln(\frac{B_\phi}{B}) - \psi(\frac{1}{2}+\frac{B_{\phi}}{B}) + ln(\frac{B_{SO}}{B}) - \psi(\frac{1}{2} + \frac{B_{SO}}{B})]
\end{equation}

where, B is the applied magnetic field, B$_{\phi}$ ($\hbar$/4el$^2_\phi$) and B$_{SO}$ ($\hbar$/4el$^2_{SO}$) are two characterstic magnetic fields related to phase coherence length (l$_\phi$) and spin-precession length (l$_{SO}$) and $\psi$ is the digamma function. The ILP theory was developed for the diffusive regime B$\textless$ $\hbar$/2el$^2_m$; where l$_m$ is the mean free path of the carriers.\cite{Nakamura2009,Iordanskii1994} For the present sample, $\hbar$/2el$^2_m$ is estimated to be 0.5 T. Figure 3(a) shows the magnetoconductance data of EuO-KTO taken at 1.8 K along with the fit using equation 3 (black line). The value of B$_{SO}$ obtained from the fit was $\backsim$2.3 T corresponding to a spin-precession length of 9 nm. Phase coherence length of 176 nm and magnetic field strength corresponding to inelastic scattering B$_\phi$=5x10$^{-2}$ T was obtained for our system. We could also calculate the value of Rashba strength parameter ($\alpha$ = (($\hbar^3$ eB$_{SO}$)/m$^{*2}$)$^{1/2}$) \cite{Narayanapillai2014} from this WAL-fitting 8.6x10$^{-12}$ eVm which is fairly high for an all-oxide systems.\cite{Veit2018,Caviglia2010,Herranz2015,Gopinadhan2015}

We have already demonstrated the existence of non-trivial Berry phase in EuO-KTO from the analysis of SdH oscillations. Non-trivial Berry phase is usually realized for the charge carriers having reciprocal space cyclotron orbits enclosing a Dirac point.\cite{Murakawa2013} Such Dirac points are usually observed in topological materials, however systems described by the Rashba Hamiltonian can also have a Dirac point and may provide an alternative source to realize Berry phase.\cite{Murakawa2013,Veit2018} The presence of strong SOC and the existence of Rashba type spin-split bands had already been reported for KTO through angle resolved photoemission spectroscopy measurements.\cite{King2012} Our observation of WAL in MR (Fig. 3(a)), non-trivial Berry phase as well as existence of two electronic Fermi surfaces is endowed with the existence of Rashba spin-split bands in the 2DEG at the interface of EuO-KTO. Based on our results of Hall and MR (SdH oscillation) data, we drew a possible band structure of this 2DEG (in Fig. 3(b)). Our magneto-transport measurements reveal that there are three bands near the Fermi level contributing towards conduction with carrier density 5x10$^{13}$ (Hall), 2.3x10$^{12}$ (SdH: OFS) and 9.8x10$^{11}$ cm$^{-2}$ (SdH: IFS). The Fermi vector calculated for the carrier density 5x10$^{13}$ is K$_{f1}$ (2n$\pi$)$^{1/2}$ = 1.8x10$^9$ m$^{-1}$. Using this K$_{f1}$, Fermi energy was calculated to be 0.36 eV by taking the effective mass value of 0.3 m$_e$ (E$_f$=($\hbar^2$k$_{f1}^2$)/(2m$^*$)).\cite{Cooper2012} Then we drew the energy dispersion relation using E=($\hbar^2$k$^2$)/(2m$^*$) and marked the corresponding Fermi level (E$_F$=0.36 eV). This is our outermost band and labelled as 1 in Fig. 3(b). The K$_{f1}$ value obtained in our case (1.8x10$^9$ m$^{-1}$) is in agreement with the previous reports on KTO (2x10$^9$ m$^{-1}$). \cite{King2012}
Then for other two types of carriers with density 2.3x10$^{12}$ and 9.8x10$^{11}$ cm$^{-2}$ (estimated from the SdH  oscillation), we calculated the respective K$_f$ (labelled as K$_{f2}$ and K$_{f3}$ respectively). The calculated values of K$_{f2}$ and K$_{f3}$ for OFS and IFS were 5.3x10$^8$ and 3.6x10$^8$ m$^{-1}$ respectively (Fig. 3(c)). Then by using the expression of Rashba energy eigenvalues and equating K$_{f2}$+K$_o$ = K$_{f3}$-K$_o$,which ensures that the OFS and IFS are at same energy level, we calculated K$_o$ to be $\sim$0.01 $\AA^{-1}$. Next we generated the Rashba spin-split energy bands by using the dispersion relation:\cite{Xiang2015}

\begin{equation}
E\textpm = \frac{\hbar^2}{2m^*} (k+k_o)^2
\end{equation}

These two energy bands were then shifted to satisfy the condition that energy at K$_{f1}$, K$_{f2}$ and K$_{f3}$ is same and equal to E$_F$ (0.36eV).
Alternatively, the value of K$_o$ can also be determined by estimating the characteristic magnetic field related to the spin precession length from the fitting of out-of-plane MR data by ILP theory.\cite{Iordanskii1994} The value of K$_o$ calculated from ILP fitting was 0.012 $\AA^{-1}$ which is fairly in agreement with the one calculated from SdH analysis (0.01 $\AA^{-1}$).  Also, it was found that the $\Delta_E$ value between the upper (2 or/and 3) and lower bands (band 1) is 0.35 eV, a similiar number is reported in previous ARPES measurements.\cite{King2012} The agreement of the k$_o$ values from SdH analysis and WAL fitting data further strengthen our analysis of band diagram.

\begin{figure}[t!] \scalebox{1}{\includegraphics{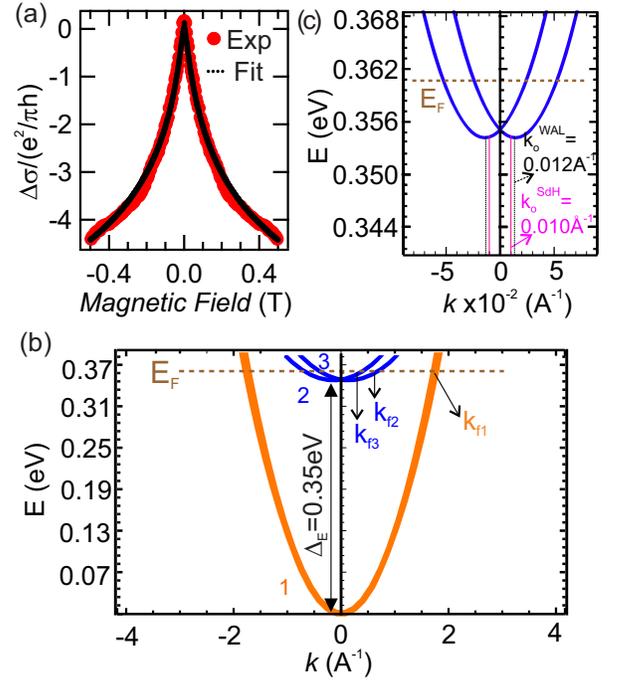}}
\caption{(Color online) (a) Magnetoconductance plot of EuO-KTO sample as a function of magnetic field showing WAL due to high SOC. (b) Possible band diagram near the Fermi level deduced from the magnetotransport measurements. (c) Zoomed region of bands 2 and 3 near Fermi level.}
\end{figure}

\begin{figure}[t!] \scalebox{1.1}{\includegraphics{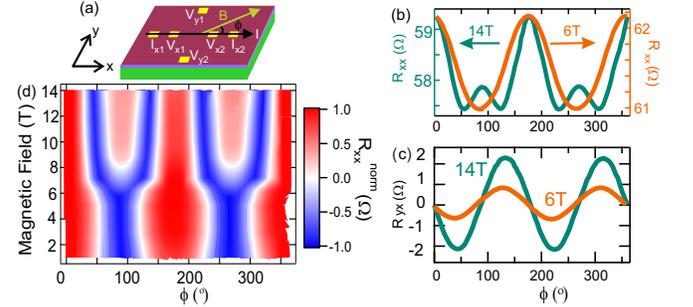}}
\caption{(Color online) (a) Schematic of the measurement geometry. (b) and (c) R$_{xx}$ and R$_{yx}$ as a function of scan angle at magnetic field of 14 (Left axis) and 6 T (Right axis) respectively. (d) Applied magnetic field and angle dependent contour plot for normalized R$_{xx}$.} 
\end{figure}

To check for topological character of the carriers arising from the Dirac point of the Rashba hamiltonian, we performed in-plane MR and Hall measurements. In these measurements, we applied magnetic field in the plane of the sample and measured longitudinal (R$_{xx}$) and transverse (R$_{yx}$) magnetoresistance by varying the angle ($\phi$) between the applied magnetic field and current (I) as shown in schematic of Fig. 4(a).\cite{Veit2018,Taskin2017,Rakhmilevich2018}  Figure 4(b) shows the measured R$_{xx}$ for applied magnetic field of 14 and 6 T. It can be seen that on varying the angle between the applied field and current, the sample shows anisotropic magnetoresistance (AMR) which changes from two fold oscillations to four fold when the magnitude of magnetic field is increased. Also, the transverse magnetoresistance known as planar Hall effect (PHE) was found to show oscillatory behavior (Fig. 4(c)). However, in this case, only the amplitude of oscillations decrease but the nature of the oscillations remains same. Figure 4(d) shows the contour plot of the normalized AMR obtained on varying the angle $\phi$ and the magnitude of applied magnetic field. The normalized resistance R$_{xx}^{norm}$ = (R - R$_{symm}$)/R$_o$. Where, R$_{symm}$ = R$_{min}$ + (R$_{max}$ - R$_{min}$)/2 and  R$_{min}$ is minimum value of R$_{xx}$,  R$_{max}$ is maximum value of  R$_{xx}$ and R$_{o}$ is the value of R$_{xx}$ at 0$^o$. This type of AMR is observed generally for topological systems and was also predicted for Rashba-type systems.\cite{Trushin2009,Kozlov2019} But this is worth noting that this is a unique system having ingredient of strong SOC, relativistic conducting electron, symmetry breaking and ferromagnetism. Hence, the observed AMR and PHE demand an elaborated theoretical study.

In conclusion, through the analysis of SdH oscillation we found the presence of two Fermi surfaces for the conducting interface of EuO-KTO containing non-trivial Berry phases. This suggests that both the Fermi surfaces are enclosing Dirac point in reciprocal space. In addition, observation of WAL, PHE and AMR is not only endowed with a non-zero SOC, but is also expected to exhibit a prominent Rashba effect in this system. Our calculation of band dispersion from the analysis of transport data suggests the presence of a significant Rashba spin-split bands. Also, we would like to point out that, the presence of non-trivial Berry phase and PHE are very similar to that observed for Dirac/Weyl Fermions in topological materials. Our observations suggest that perovskite oxides with strong SOC and relativistic conduction electron could be a hunting ground not only spintronic materials but also for emergent physics.

\section{ACKNOWLEDGMENTS}
Authors acknowledge financial support from DST Nano Mission project number (SR/NM/NS-1007/2015).

\end{document}